\documentclass[useAMS,usenatbib]{mn2e}
\usepackage{txfonts}
\usepackage{natbib}
\usepackage[all]{xy}
\usepackage[dvips]{graphicx}

\def\del#1{{}}

\newcommand{\ltsima}{$\; \buildrel < \over \sim \;$}
\newcommand{\lsim}{\lower.5ex\hbox{\ltsima}}
\newcommand{\gtsima}{$\; \buildrel > \over \sim \;$}
\newcommand{\gsim}{\lower.5ex\hbox{\gtsima}}
\newcommand{\bra}{\langle}
\newcommand{\ket}{\rangle}

\newcommand{\dd}{\mathrm{d}}

\newcommand{\ko}{{{\bf k}_1}}
\newcommand{\kt}{{{\bf k}_2}}
\newcommand{\kth}{{{\bf k}_3}}

\title[weak lensing and non-Gaussianities]
{A weak lensing view on primordial non-Gaussianities}
\author[B.M. Sch{\"a}fer, A. Grassi, M. Gerstenlauer, C.T. Byrnes]{Bj{\"o}rn Malte Sch{\"a}fer$^1$\thanks{e-mail: bjoern.malte.schaefer@uni-heidelberg.de}, Alessandra Grassi$^1$, Mischa Gerstenlauer$^2$ and Christian T. Byrnes$^3$\\
$^1$Astronomisches Recheninstitut, Zentrum f{\"u}r Astronomie, Universit{\"a}t Heidelberg, M{\"o}nchhofstra{\ss}e 12, 69120 Heidelberg, Germany\\
$^2$Institut f{\"u}r theoretische Physik, Universit{\"a}t Heidelberg, Philosophenweg 16, 69120 Heidelberg, Germany\\
$^3$Fakult{\"a}t f{\"u}r Physik, Universit{\"a}t Bielefeld, Postfach 100131, 33501 Bielefeld, Germany}

\begin{document}
\pagerange{\pageref{firstpage}--\pageref{lastpage}}
\pubyear{2011}
\maketitle
\label{firstpage}

\begin{abstract}
We investigate the signature of primordial non-Gaussianities in the weak lensing bispectrum, in particular the signals generated by local, orthogonal and equilateral non-Gaussianities. The questions we address include the signal-to-noise ratio generated in the Euclid weak lensing survey (we find the $1\sigma$-errors for $f_\mathrm{NL}$ are 200, 575 and 1628 for local, orthogonal and equilateral non-Gaussianities, respectively), misestimations of $f_\mathrm{NL}$ if one chooses the wrong non-Gaussianity model (misestimations by up to a factor of $\pm3$ in $f_\mathrm{NL}$ are possible, depending on the choice of the model), the probability of noticing such a mistake (improbably large values for the $\chi^2$-functional occur from $f_\mathrm{NL}\sim200$ on), degeneracies of the primordial bispectrum with other cosmological parameters (only the matter density $\Omega_m$ plays a significant role), and the subtraction of the much larger, structure-formation generated bispectrum. If a prior on a standard $w$CDM-parameter set is available from Euclid and Planck, the structure formation bispectrum can be predicted accurately enough for subtraction, and any residual structure formation bispectrum would influence the estimation of $f_\mathrm{NL}$ to a minor degree. Configuration-space integrations which appear in the evaluation of $\chi^2$-functionals and related quantities can be carried out very efficiently with Monte-Carlo techniques, which reduce the complexity by a factor of $\mathcal{O}(10^4)$ while delivering sub-percent accuracies. Weak lensing probes smaller scales than the CMB and hence provide an additional constraint on non-Gaussianities, even though they are not as sensitive to primordial non-Gaussianities as the CMB.
\end{abstract}


\section{Introduction}
As cosmological data improves, it is becoming increasingly feasible to probe models of the early universe. In particular, primordial non-Gaussianity has emerged as a leading window onto the physics of inflation and the early universe \citep[for reviews, see][]{2004PhR...402..103B, Komatsu:2010hc, 2011arXiv1102.5052L}. Although non-Gaussian perturbations could in principle take any form, in practice searching for just a few shapes of the bispectrum (3-point correlation function) allows one to discriminate between entire classes of models \citep{2003PhRvD..67l1301B, Komatsu:2009kd, 2011ApJS..192...18K}. Many alternative models to inflation produce the same non-Gaussian shapes, and therefore can also be constrained at the same time. 

The theory of primordial non-Gaussianities is very evolved, and predictions for non-Gaussianities in different observational channels have been made: number counts and large-scale structure statistics \citep{2010AdAst2010E..64V, 2010CQGra..27l4011D, 2010AdAst2010E..89D, 2011MNRAS.414.1545F}, the cosmic microwave background even outside the Sachs-Wolfe regime \citep{2007PhRvD..76h3523F, 2010PhRvD..82b3502F} and gravitational lensing \citep{2011arXiv1103.5396F, 2011MNRAS.411..595P, 2011ApJ...728L..13M, 2011arXiv1104.0926J}.

Most constraints on non-Gaussianities are reported using CMB-observations, either by measuring the bispectrum of the temperature perturbation directly \citep{2003NewAR..47..797K, 2006JCAP...05..004C, 2008PhRvL.100r1301Y, 2009ApJ...706..399C, 2009ApJS..180..330K, 2009MNRAS.397..837V, 2010MNRAS.404..895V, 2011MNRAS.411.2019C}, by measuring the skewness of weighted averaged CMB-patches \citep{2004ApJ...613...51M} or by quantifying the corresponding Minkowski functionals \citep{2005MNRAS.358..684C, 2007MNRAS.377.1668G, 2008MNRAS.389.1439H}. The tightest bounds on the amplitude $f_\mathrm{NL}$ of the bispectrum, $-10\leq f_\mathrm{NL}\leq74$ has been obtained by \citet{2011ApJS..192...18K}.

So far, observational constraints on the weak lensing bispectrum \citep{2003A&A...397..405B, 2011arXiv1105.2309S} mainly concerned the non-Gaussianities generated by structure formation and their breaking of the $\Omega_m$-$\sigma_8$ degeneracy \citep{2003A&A...409..411M}, and first results have been obtained using the skewness of the aperture-mass statistic \citep{1998MNRAS.296..873S, 2011MNRAS.410..143S}.

In this paper we forecast the constraints which the weak lensing bispectrum will be able to make on primordial non-Gaussianity, especially with a view to Euclid. Although the constraints are not competitive with the CMB if the non-Gaussianities are scale independent, they are complementary since they probe smaller scales compared to the CMB and provide constraints on a possible scale-dependence \citep{2008JCAP...04..014L}. A scale dependence of $f_\mathrm{NL}$ at the same order as the spectral index of the power spectrum is natural \citep{2010JCAP...10..004B} and it may be much stronger, e.g.~a ``step--function" which is zero on large scales and large on small scales \citep{Riotto:2010nh}. Weak lensing also has lower systematic errors than other large scale structure probes such as the galaxy bispectrum, scale dependent bias and cluster counts. Because the weak shear provides a linear mapping of the cosmic matter distribution, the statistical properties of the source field are conserved in the observable.

After a brief recapitulation of cosmology and structure formation in Sect.~\ref{sect_cosmology} we introduce primordial  and structure formation non-Gaussianities in Sect.~\ref{sect_nong}, in particular the bispectral shapes and the motivation for studying them. The mapping of non-Gaussianities by weak gravitational lensing is treated in Sect.~\ref{sect_weak_lensing}, where we investigate the properties of the weak lensing bispectrum in its scale- and configuration dependence, and how it builds up as a function of survey depth. Statistical questions concerning the signal strength and misestimations of the non-Gaussianity parameter are addressed in Sect.~\ref{sect_statistics}, before we focus on systematic errors in the non-Gaussianity parameter due to incompletely removed structure formation non-Gaussianities in Sect.~\ref{sect_systematics}. We summarise our main results in Sect.~\ref{sect_summary} and provide visualisations of the weak lensing bispectrum sourced by different non-Gaussianity shapes in Appendix~\ref{sect_configuration}.

The reference cosmological model used is a spatially flat $w$CDM cosmology with Gaussian adiabatic initial perturbations for the cold dark matter density. The specific parameter choices are $\Omega_m = 0.25$, $n_s = 1$, $\sigma_8 = 0.8$, $\Omega_b=0.04$ and $H_0=100\: h\:\mathrm{km}/\mathrm{s}/\mathrm{Mpc}$, with $h=0.72$. The dark energy equation of state is constant in time with a value of $w=-0.9$.

\section{Cosmology and structure formation}\label{sect_cosmology}
In spatially flat dark energy cosmologies with the matter density parameter $\Omega_m$, the Hubble function $H(a)=\dd\ln a/\dd t$ is given by
\begin{equation}
\frac{H^2(a)}{H_0^2} = \frac{\Omega_m}{a^{3}} + \frac{1-\Omega_m}{a^{3(1+w)}},
\end{equation}
for a constant dark energy equation of state-parameter $w$. Comoving distance $\chi$ and scale factor $a$ are related by
\begin{equation}
\chi = c\int_a^1\:\frac{\dd a}{a^2 H(a)},
\end{equation}
such that the comoving distance is given in units of the Hubble distance $\chi_H=c/H_0$. For the linear matter power spectrum $P(k)$ which describes the Gaussian fluctuation properties of the linearly evolving density field $\delta$,
\begin{equation}
\bra \delta(\bmath{k})\delta(\bmath{k}^\prime)\ket = (2\pi)^3\delta_D(\bmath{k}+\bmath{k}^\prime)P(k)
\end{equation}
the ansatz $P(k)\propto k^{n_s} T^2(k)$ is chosen with the transfer function $T(k)$, which is well approximated by the fitting formula
\begin{equation}
T(q) = \frac{\ln(1+2.34q)}{2.34q}\times p(q)^{-1/4},
\end{equation}
for low-matter density cosmologies \citep{1986ApJ...304...15B}. The polynomial $p(q)$ is given by $p(q) = 1+3.89q+(16.1q)^2+(5.46q)^3+(6.71q)^4$. The wave vector $k=q\Gamma$ enters rescaled by the shape parameter $\Gamma$ \citep{1995ApJS..100..281S},
\begin{equation}
\Gamma=
\Omega_m h\exp\left[-\Omega_b\left(1+\frac{\sqrt{2h}}{\Omega_m}\right)\right].
\end{equation}
The fluctuation amplitude is normalised to the value $\sigma_8$ on the scale $R=8~\mathrm{Mpc}/h$,
\begin{equation}
\sigma_R^2 = \frac{1}{2\pi^2}\int\dd k\: k^2 W^2_R(k) P(k),
\end{equation}
with a Fourier-transformed spherical top-hat $W_R(k)=3j_1(kR)/(kR)$ as the filter function. $j_\ell(x)$ denotes the spherical Bessel function of the first kind of order $\ell$ \citep{1972hmf..book.....A}. The linear  growth of the density field, $\delta(\bmath{x},a)=D_+(a)\delta(\bmath{x},a=1)$, is described by the growth function $D_+(a)$, which is the solution to the growth equation \citep{1997PhRvD..56.4439T, 1998ApJ...508..483W, 2003MNRAS.346..573L},
\begin{equation}
\frac{\dd^2}{\dd a^2}D_+(a) + \frac{1}{a}\left(3+\frac{\dd\ln H}{\dd\ln a}\right)\frac{\dd}{\dd a}D_+(a) = 
\frac{3}{2a^2}\Omega_m(a) D_+(a).
\label{eqn_growth}
\end{equation}

\section{non-Gaussianities}\label{sect_nong}

\subsection{Primordial non-Gaussianities}\label{sect_inflation_bispectrum}
We write the primordial bispectra in terms of the Bardeen curvature perturbation $\Phi$ \citep{Bardeen:1980kt, Bardeen:1983qw}, which may be related to the primordial curvature perturbation $\zeta=5\Phi/3$ \citep[see e.g. chapter 8 in][]{Riotto:2002yw} and the CMB temperature anisotropy in the Sachs-Wolfe limit $\Delta T/T=-\Phi/3$ \citep{Sachs:1967er}. The bispectrum of $\Phi$ is defined by
\begin{eqnarray} \langle\Phi(\ko)\Phi(\kt)\Phi(\kth)\rangle = 
(2\pi)^3\delta_{D}(\ko+\kt+\kth)B_\Phi(k_1,k_2,k_3). 
\end{eqnarray}
We will be particularly interested in three bispectral shapes, which cover the expected shape from a wide range of inflationary models \citep{Komatsu:2010hc}. They are defined as:
\begin{enumerate}
\item{ {\bf Local shape.} This is defined by 
\begin{eqnarray} 
B^{local}_\Phi(k_1,k_2,k_3)=
2A^2 f_\mathrm{NL}^{local}\left(k_1^{n_s-4}k_2^{n_s-4} + (2\;perm)\right), 
\end{eqnarray}
where the amplitude $A$ is defined by $P_{\Phi}(k)=Ak^{n_s-1}$. It may arise through a simple Taylor expansion about the Gaussian (linearised) perturbation $\Phi(x)=\Phi_G(x)+f_\mathrm{NL}\Phi_G^2$, although this is not the most general ansatz for $\Phi$ which gives rise to the local bispectrum. The local shape typically arises from super-horizon evolution of the curvature perturbation ($k\ll aH$), which occurs for example in some multifield inflation models \citep{Byrnes:2008wi,Elliston:2011dr}, during modulated reheating \citep{Dvali:2003em}, in the curvaton scenario \citep{Lyth:2002my} (the last three models are closely connected \citep{Alabidi:2010ba}), as well as the ekpyrotic scenario \citep{Lehners:2010fy} and non-local inflation \citep{Barnaby:2008fk}.}
\item{ The {\bf equilateral shape} is given by
\begin{eqnarray} 
B^{equil}_\Phi(k_1,k_2,k_3)
& = & 
6A^2 f_\mathrm{NL}^{equil} \left(-2\left(k_1k_2k_3\right)^{2(n_s-4)/3}\right.\nonumber\\
&&  -  
\left[ \left(k_1k_2\right)^{(n_s-4)} +(2\;perm)\right] \nonumber\\ 
&& +  
\left.\left[k_1^{(n_s-4)/3}k_2^{2(n_s-4)/3}k_3^{n_s-4} +(5\;perm) \right] \right).  
\end{eqnarray}
This shape typically arises in models with non-canonical kinetic terms, the most studied example being Dirac-Born-Infeld inflation \citep{Silverstein:2003hf, Alishahiha:2004eh}. Also various other models can produce this shape \citep{ArkaniHamed:2003uz, Seery:2005wm, Chen:2006nt, Cheung:2007st, Li:2008qc}.}
\item{The {\bf orthogonal shape} is given by 
\begin{eqnarray} 
B^{ortho}_\Phi(k_1,k_2,k_3)
& = & 
6A^2 f_\mathrm{NL}^{ortho} \left(-8\left(k_1k_2k_3\right)^{2(n_s-4)/3}\right.\nonumber\\ 
&& -  
3 \left[ \left(k_1k_2\right)^{(n_s-4)} +(2\;perm)\right] \nonumber\\ 
&& +  
3 \left.\left[k_1^{(n_s-4)/3}k_2^{2(n_s-4)/3}k_3^{n_s-4} +(5\;perm) \right] \right)\,,  
\end{eqnarray}
\citep{Senatore:2009gt} and it was constructed in order to be orthogonal to both the equilateral shape and to a lesser extent the local shape.}
\end{enumerate}
The local model is maximised for squeezed triangles $k_1\ll k_2\simeq k_3$, the equilateral model is maximised for equilateral triangles $k_1\simeq k_2\simeq k_3$ while the orthogonal model receives contributions from a broader range of triangles. Another frequently considered shape is the enfolded one, which is maximised for ``flattened" isosceles triangles $k_1\simeq k_2 \simeq k_3/2$, but this can be written as a linear combination of the three shapes above. We note that although the above three shapes cover many classes of non-Gaussian models, there do exist other shapes which cannot be written as a combination of the above three shapes including localised or oscillating bispectra, which may be caused by a feature in the inflatons potential \citep{Chen:2006xjb,2011arXiv1106.5384A}, particle production while observable modes are crossing the horizon \citep{Barnaby:2010sq} \citep[but a burst of particle production later in inflation generates local non-Gaussianity,][]{Battefeld:2011yj}, or an inflaton potential with superimposed oscillations \citep{Chen:2008wn,Chen:2011zf}. 

Because $f_\mathrm{NL}$ for all three shapes is normalised to an equilateral triangle, but the signal-to-noise is maximised for different triangle shapes depending on the configuration it is not surprising that the error bars on the three shapes are significantly different \citep{Fergusson:2008ra}. In line with this expectation, we find that weak lensing can constrain the local model most tightly and the equilateral model least well. When investigating relative magnitudes between structure formation bispectra and primordial ones, we restrict ourselves to the equilateral case, as all primordial bispectra have equal values for this configuration. An alternative normalisation was proposed by \citet{Fergusson:2008ra}.

A mild scale dependence of $f_\mathrm{NL}$ is natural, for both the local model \citep{2010JCAP...10..004B,Byrnes:2010xd,Huang:2010cy}, and the equilateral and orthogonal models \citep{Chen:2005fe,Bartolo:2010im,Noller:2011hd,Burrage:2011hd}, and it may be much stronger, e.g.~a ``step--function" which is zero on large scales and large on small scales \citep{Riotto:2010nh}. Although we treat $f_\mathrm{NL}$ as constant in this paper, the motivation for considering scale-dependence is important because weak lensing probes smaller scales than the CMB and hence the CMB bounds may not apply here, see Sec.~\ref{sect_weak_lensing}. Observational probes have been considered in \citep{2008JCAP...04..014L,Sefusatti:2009xu,Shandera:2010ei,Becker:2010hx}. 

For converting the bispectrum of the potential fluctuations to those of the density field we use the Newtonian Poisson equation for each occurence of the potential in the bispectrum \citep{2011arXiv1103.1876M},
\begin{equation}
\Delta\Phi = \frac{3}{2}\frac{\Omega_m}{\chi_H^2}\delta
\rightarrow
\delta(k,a) = \frac{2}{3\Omega_m} D_+(a) (\chi_H k)^2 T(k) \:\Phi(k)
\end{equation}
The horizon entry of each mode is governed by the transfer function $T(k)$ and it grows $\propto D_+(a)$ in the linear regime, such that
\begin{equation}
B_\delta(k_1,k_2,k_3,a) = 
\prod_{i=1}^3 \left(\frac{2}{3\Omega_m} D_+(a) (\chi_H k_i)^2 T(k_i)\right)\: B_\Phi(k_1,k_2,k_3).
\end{equation}
We choose the normalisation factor $A$ to be consistent for each linearly evolving mode of the density field with our definition of $\sigma_8$.

\subsection{Non-Gaussianities from structure formation}\label{sect_sf_bispectrum}
Nonlinear processes in structure formation break the homogeneity of the growth equation and generate non-Gaussian features in the initially close to Gaussian density field. From Eulerian perturbation theory \citep[see][]{1995PhR...262....1S, 2002PhR...367....1B, 2011PhRvD..83h3518M, 2001MNRAS.325.1312S}, the first order contribution to the bispectrum $B_\delta(\bmath{k}_1,\bmath{k}_2,\bmath{k}_3)$ \citep[for an introduction, see ][]{1984ApJ...277L...5F, 1984ApJ...279..499F} of the density field from nonlinear structure formation is given by:
\begin{equation}
B_\delta(\bmath{k}_1,\bmath{k}_2,\bmath{k}_3,a) = 
\sum_{{i,j=1,2,3\atop i\neq j}} D_+^4(a)\:M(\bmath{k}_i,\bmath{k}_j)\:P(k_i)P(k_j),
\end{equation}
where the classical mode coupling function is $M(\bmath{k}_i,\bmath{k}_j)$
\begin{equation}
M(\bmath{k}_i,\bmath{k}_j) = \frac{10}{7} +
\left(\frac{k_i}{k_j}+\frac{k_j}{k_i}\right)x + \frac{4}{7}x^2.
\end{equation}
$x=\bmath{k}_i\bmath{k}_j/(k_ik_j)$ denotes the cosine between the wave vectors $\bmath{k}_i$ and $\bmath{k}_j$. Due to the fact that $P(k,a)$ grows $\propto D_+^2(a)$ in linear structure formation, the bispectrum scales with $D_+^4(a)$ in lowest order perturbation theory. In terms of non-Gaussianity parameters and configuration dependences, structure formation non-Gaussianities are strongest for the squeezed configuration because the mode coupling function $M(\bmath{k}_i,\bmath{k}_j)$ assumes the largest values for parallel wave vectors (with the cosine being one, $x=1$), and therefore resembles non-Gaussianities of the local type. Their strength in the weak shear bispectrum corresponds to an $f_\mathrm{NL}$-parameter of $\mathcal{O}(10^4)$, i.e. two orders of magnitude larger than the primordial non-Gaussianities weak lensing can probe.

\section{Weak gravitational lensing}\label{sect_weak_lensing}

\subsection{Convergence spectrum}
The weak lensing convergence $\kappa$ follows from a line-of-sight integration weighted with the lensing efficiency $W_\kappa(\chi)$ \citep[for reviews, see ][]{2001PhR...340..291B, 2010CQGra..27w3001B},
\begin{equation}
\kappa = \int_0^{\chi_H}\dd\chi\: W_\kappa(\chi)\delta
\end{equation}
and reflects, because of its linearity, all statistical properties of the density field $\delta$. The weak lensing efficiency is given by
\begin{equation}
W_\kappa(\chi) = \frac{3\Omega_m}{2a}\frac{1}{\chi_H^2}G(\chi)\chi,
\end{equation}
with the weighted distance distribution $G(\chi)$ of the lensed galaxies,
\begin{equation}
G(\chi) = 
\int_\chi^{\chi_H}\dd\chi^\prime\:q(z)\frac{\dd z}{\dd\chi^\prime}\frac{\chi^\prime-\chi}{\chi^\prime}.
\end{equation}
The spectrum $C_\kappa(\ell)$ then results from applying Limber's equation \citep{1954ApJ...119..655L},
\begin{equation}
C_\kappa(\ell) = \int_0^{\chi_H}\frac{\dd\chi}{\chi^2}\:W_\kappa^2(\chi) P(k=\ell/\chi,a).
\end{equation}
For the galaxy redshift distribution $q(z)$ we assume a standard shape,
\begin{equation}
q(z) = q_0\left(\frac{z}{z_0}\right)^2\exp\left(-\left(\frac{z}{z_0}\right)^\beta\right)\dd z
\quad\mathrm{with}\quad \frac{1}{q_0}=\frac{z_0}{\beta}\Gamma\left(\frac{3}{\beta}\right),
\end{equation}
with the median redshift set to 0.9, as projected for Euclid.

\subsection{Convergence bispectrum}
Similarly as in the case of the weak shear spectrum $C_\kappa(\ell)$ we use the Limber-equation in the flat-sky approximation,
\begin{equation}
B_\kappa(\bmath{\ell}_1,\bmath{\ell}_2,\bmath{\ell}_3) =
\int_0^{\chi_H}\frac{\dd\chi}{\chi^4}\:W^3(\chi)\:
B_\delta(\bmath{k}_1,\bmath{k}_2,\bmath{k}_3,a),
\label{eqn_lensing_bispectrum}
\end{equation}
with $\bmath{k}_p = \bmath{\ell}_p/\chi$, $p=1,2,3$, for projection of the flat-sky convergence bispectrum $B_\kappa$ \citep{1998MNRAS.296..873S, 1999ApJ...522L..21H,2003MNRAS.340..580T,2003MNRAS.344..857T, 2004MNRAS.348..897T, 2005PhRvD..72h3001D}. The spherical bispectrum $B_\kappa(\ell_1,\ell_2,\ell_3)$ is related to the flat-sky bispectrum $B_\kappa(\bmath{\ell}_1,\bmath{\ell}_2,\bmath{\ell}_3)$ by \citep{1991ApJ...380....1M, 1992ApJ...388..272K}
\begin{equation}
B_\kappa(\ell_1,\ell_2,\ell_3) \simeq \left(
\begin{array}{ccc}
\ell_1 & \ell_2 & \ell_3\\
0 & 0 & 0
\end{array}
\right) \sqrt{\frac{\prod_{p=1}^3(2\ell_p+1)}{4\pi}}
B_\kappa(\bmath{\ell}_1,\bmath{\ell}_2,\bmath{\ell}_3),
\end{equation}
where
\begin{equation}
\left(\begin{array}{ccc}\ell_1 & \ell_2 & \ell_3\\ 0 & 0 &
0\end{array}\right)^2 = \frac{1}{2}\int_{-1}^{+1}\dd x\:
P_{\ell_1}(x)P_{\ell_2}(x)P_{\ell_3}(x),
\end{equation}
denotes the Wigner-$3j$ symbol, which results from integrating over three Legendre polynomials $P_\ell(x)\,$ ($x=\cos\theta$). The Wigner-$3j$ symbol nulls configurations which would violate the triangle inequality, $\left|\ell_i-\ell_j\right|\leq \ell_k \leq \left|\ell_i+\ell_j\right|$ \citep{1972hmf..book.....A}. The factorials in the Wigner-$3j$ symbol are evaluated using the Stirling approximation for the $\Gamma$-function, 
\begin{equation}
\Gamma(n+1)=n!
\quad\mathrm{with}\quad
\Gamma(x)\simeq\sqrt{2\pi}\:\exp(-x)\:x^{x-\frac{1}{2}},
\end{equation}
for $x\gg 1$ \citep{1972hmf..book.....A}. At this point, it is appropriate to recall two important issues related to the weak shear bispectrum as a line of sight-integrated quantity: The line-of-sight integration causes the non-Gaussianities in the convergence to be weaker than that of the source field, as a consequence of the central limit theorem, because many uncorrelated lensing effects \citep[if the Born approximation is invoked and lens-lens coupling is neglected, see][]{2002ApJ...574...19C, 2006JCAP...03..007S, 2010A&A...523A..28K, 2011arXiv1101.4769S} add up to the signal \citep{2011arXiv1104.0926J}. Secondly, the evaluation of the wave vector $k=\ell/\chi$ in the source field bispectrum $B_\delta$ generates a mixing of scales when the distance $\chi$ runs over the integration range such that the observed weak lensing bispectrum is a superposition of density field bispectra of varying scale and fixed projected configuration.

\subsection{Properties of the weak lensing bispectrum}
Fig.~\ref{fig_config} gives a 3-dimensional visual impression of the three different bispectra as observed by weak shear. The weak shear bispectra are given as dimensionless bispectra, by multiplication with the prefactor $(\ell_1\ell_2\ell_3)^{4/3}\sim\ell^4$. There are clear differences in the configuration dependence: Local non-Gaussianities provide large amplitudes for squeezed configurations, i.e. in the corners of the domain admissible by the triangle inequality, orthogonal non-Gaussianities are largest for folded configurations and the equilateral bispectra assume large values if the three multipole orders are equal.

\begin{figure}
\begin{center}
\resizebox{\hsize}{!}{\includegraphics{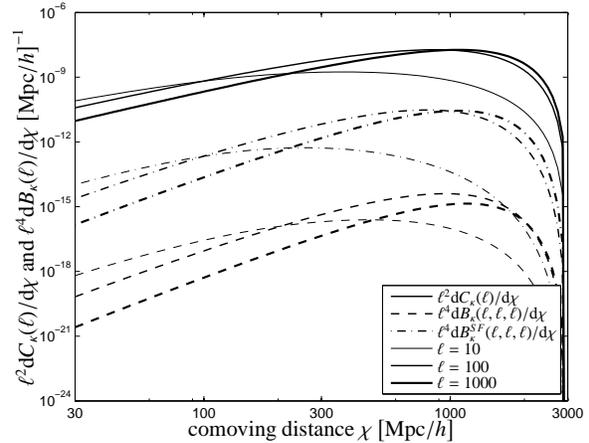}}
\end{center}
\caption{Contributions $\ell^2 \dd C_\kappa(\ell)/\dd\chi$ to the lensing spectrum (solid line), and $\ell^4 \dd B_\kappa(\ell,\ell,\ell)/\dd\chi$ to the lensing bispectra for both primordial non-Gaussianities (dashed line, $f_\mathrm{NL}=1$) and structure formation non-Gaussianities (dash-dotted line), for $\ell=10$ (thin lines), $\ell=100$ (medium lines) and $\ell=1000$ (thick lines). The bispectra are plotted for the equilateral configuration.}
\label{fig_contribution}
\end{figure}

Fig.~\ref{fig_contribution} illustrates the contribution $\dd C_\kappa(\ell)/\dd\chi$ to the spectrum and the contribution $\dd B_\kappa(\ell,\ell,\ell)/\dd\chi$ to the equilateral bispectrum as a function of comoving distance. For the relevant range of multipoles, modes with wave numbers in the range $0.1\ldots1~\mathrm{Mpc}/h$ are being probed by weak shear, which are larger than those wave numbers measurable in the primary CMB bispectrum and emphasises the necessity of measuring $f_\mathrm{NL}$ in its scale dependence, in a similar way as advocated by \citet{2008JCAP...04..014L} for number counts.

\begin{figure}
\begin{center}
\resizebox{\hsize}{!}{\includegraphics{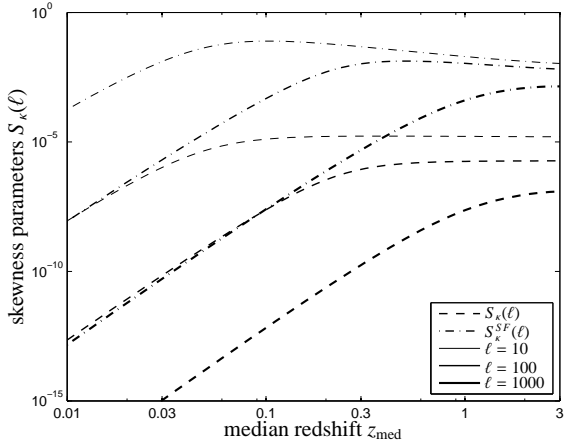}}
\end{center}
\caption{Skewness parameter $S_\kappa(\ell)$ for both primordial non-Gaussianities (dashed line, $f_\mathrm{NL}=1$) and structure formation non-Gaussianities (dash-dotted line) as a function of the survey depth, for $\ell=10$ (thin lines), $\ell=100$ (medium lines) and $\ell=1000$ (thick lines).}
\label{fig_skewness}
\end{figure}

The skewness parameters $S_\kappa(\ell)$ are defined as the ratio between the squared equilateral bispectrum and the cubed spectrum,
\begin{equation}
S_\kappa(\ell) = \sqrt{\frac{B_\kappa^2(\ell,\ell,\ell)}{C_\kappa^3(\ell)}}.
\end{equation}
$S_\kappa(\ell)$ is proportional to $f_\mathrm{NL}$ (and to $\sigma_8$ to lowest order) and is independent of $\Omega_m$. It will become relevant for the signal-to-noise ratio $\Sigma(\ell)$ (see eqn.~\ref{eqn_s2n} in Sect.~\ref{sect_s2n}). These skewness-parameters are depicted in Fig.~\ref{fig_skewness} as a function of median redshift $z_\mathrm{med}$ of the lensing survey for both the primordial and the structure formation induced weak lensing bispectrum. The skewness parameters increase with increasing survey depth and in case of primordial non-Gaussianities saturate at redshifts of unity, which is an effect of the time evolution of the gravitational potential being mapped out by weak lensing, as perturbations in the potentials decay in the dark energy-dominated phase and are constant in the matter-dominated phase. Structure formation non-Gaussianities decrease slightly for deeper reaching surveys, which is caused by the fact that the structure formation skewness's are building up during $\Omega_m$-domination, as they scale $\propto D_+^4/a^3\sim a$ in contrast to primordial non-Gaussianities in the potential, which are constant with their scaling $\propto (D_+(a)/a)^3\sim\mathrm{const}$. The plot suggests that with the redshift range probed by Euclid the largest-possible skewness's are being observed.

\section{statistics}\label{sect_statistics}

\subsection{What signal-to-noise ratio can one expect?}\label{sect_s2n}
The cumulative signal-to-noise ratio $\Sigma(\ell)$ for the weak lensing bispectrum $B_\kappa(\ell_1,\ell_2,\ell_3)$ up to multipole order $\ell$ is given by \citep{1999ApJ...522L..21H}
\begin{equation}
\Sigma^2(\ell) = 
\sum_{\ell_1=\ell_\mathrm{min}}^\ell
\sum_{\ell_2=\ell_\mathrm{min}}^\ell
\sum_{\ell_3=\ell_\mathrm{min}}^\ell \frac{B^2_\kappa(\ell_1,\ell_2,\ell_3)}{\mathrm{cov}(\ell_1,\ell_2,\ell_3)}.
\label{eqn_s2n}
\end{equation}
The summation is carried out with the condition $\ell_1\leq\ell_2\leq\ell_3$ \citep{2004MNRAS.348..897T}, such that the covariance becomes
\begin{equation}
\mathrm{cov}(\ell_1,\ell_2,\ell_3)=
\frac{\Delta(\ell_1,\ell_2,\ell_3)}{f_\mathrm{sky}}\tilde{C}(\ell_1)\tilde{C}(\ell_2)\tilde{C}(\ell_3),
\end{equation}
where the function $\Delta(\ell_1,\ell_2,\ell_3)$ counts the multiplicity of triangle configurations and is defined as
\begin{equation}
\Delta(\ell_1,\ell_2,\ell_3)= \left\{
\begin{array}
{l@{,\:}ll}
6 & & \ell_1=\ell_2=\ell_3\\
2 & & \ell_i=\ell_j\mathrm{~for~} i\neq j\\
1 & & \ell_1\neq\ell_2\neq\ell_3\neq\ell_1
\end{array}
\right.
\end{equation}
$f_\mathrm{sky}$ denotes the fraction of the observed sky and is set to $f_\mathrm{sky}=1/2$ for Euclid. The observed spectra 
\begin{equation}
\tilde{C}(\ell) = C(\ell) + \frac{\sigma^2_\epsilon}{n}\,,
\end{equation}
with the number density of ellipticity measurements per steradian $n$, which is set to 40 galaxies per squared arcminute, corresponding to the projected Euclid performance. Instead of a direct summation over $\ell_1$, $\ell_2$, $\ell_3$ we use a Monte-Carlo integration technique and consider the evaluation of eqn.~(\ref{eqn_s2n}) as a three-dimensional integration, for which we use publicly available CUBA-library \citep{2005CoPhC.168...78H}.

\begin{figure}
\begin{center}
\resizebox{\hsize}{!}{\includegraphics{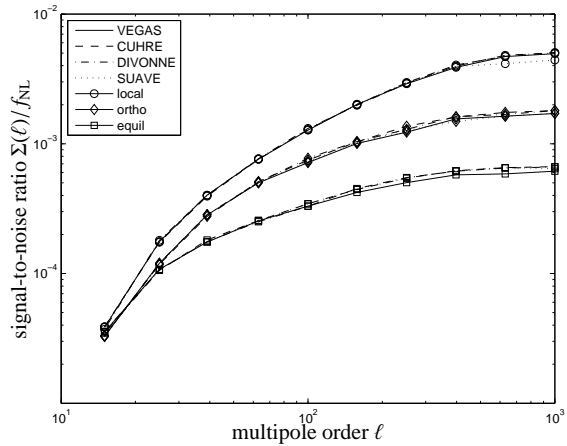}}
\end{center}
\caption{Cumulative signal-to-noise ratio $\Sigma(\ell)/f_\mathrm{NL}$ as a function of maximum multipole order $\ell$, for local (circles), orthogonal (lozenges) and equilateral (squares) non-Gaussianities. The figure compares the result from different Monte-Carlo integration routines.}
\label{fig_s2n}
\end{figure}

The cumulative signal-to-noise ratio $\Sigma(\ell)$ for a measurement of the weak shear bispectrum is depicted in Fig.~\ref{fig_s2n} as a function of $\ell$ and for all three non-Gaussianity types. As the signal strength $\Sigma$ is proportional to the non-Gaussianity parameter $f_\mathrm{NL}$, it is convenient to plot the ratio $\Sigma(\ell)/f_\mathrm{NL}$. The plot suggests that weak lensing bispectra sourced by primordial non-Gaussianities could only be measured with Euclid for $f_\mathrm{NL}$ significantly larger than 100. Orthogonal bispectra are weaker by a factor of 3 compared to local bispectra, and equilateral bispectra generate the weakest signal, being a factor of 8 weaker than local non-Gaussianities. The four different MC-integration algorithms agree well in their results for $\Sigma(\ell)$, and when the number of sampling points is chosen to be $\propto\ell$, the algorithms retain accuracies, indicating that the adaptive algorithms take account of the symmetry properties of the integrand. At $\ell=1000$, is is sufficient to compute $\mathcal{O}(10^5)$ samples, which reduces the number of evaluations of the integrand by a factor of $10^4$ compared to the exact evaluation, which would require $\mathcal{O}(10^9)$ evaluations, for precisions on the sub-percent level. We will restrict $\ell$-space integrations to multipoles $\lsim 1000$, because the increase in signal when extending the $\ell$-range is marginal for primordial non-Gaussianities and additionally, it helps to avoid scales influenced by baryonic physics and intrinsic alignments \citep{2008MNRAS.388..991S, 2011arXiv1105.1075S}.

The configuration dependence of the contribution to the integrated signal is given in Fig.~\ref{fig_config} in the appendix for the three bispectrum types considered here. The panels show the weak shear bispectrum in units of the noise, $B_\kappa(\ell_1,\ell_2,\ell_3)/\sqrt{\mathrm{cov}(\ell_1,\ell_2,\ell_3)}$ as a function of $\ell_1$, $\ell_2$ and $\ell_3$. While the local non-Gaussianity provides the largest contributions for squeezed configurations, the orthogonal non-Gaussianity shows a much uniformer contribution throughout $\ell$-space, and the equilateral non-Gaussianity is only providing significant amplitudes for very small values of $\ell$. This behavior is reflected in in the cumulative signal-to-noise ratio, as shown in Fig.~\ref{fig_s2n}. From the signal-to-noise ratios one can already estimate the accuracy for a measurement of $f_\mathrm{NL}$. The conditional Cram{\'e}r-Rao bounds $\sigma_{f_\mathrm{NL}} = 1/\sqrt{F_{f_\mathrm{NL}f_\mathrm{NL}}}$ on the non-Gaussianity parameter (which are at the same time the non-Gaussianities required to generate a signal of unity) are $\sigma_{f_\mathrm{NL}} = 200$ (local), $\sigma_{f_\mathrm{NL}} = 575$ (orthogonal) and $\sigma_{f_\mathrm{NL}} = 1628$ (equilateral) which is significantly weaker than other probes such as the primary CMB, due to the Gaussianising effect of the line-of-sight integrations \citep{2011arXiv1104.0926J}.

\subsection{Would one misestimate $f_\mathrm{NL}$ using the wrong bispectrum?}
The $\chi^2$-functional constructed for measuring the noise-weighted mismatch between the true bispectrum $B_\kappa^t$ and the wrongly assumed bispectrum $B_\kappa^w$ for interpreting the data reads:
\begin{equation}
\chi^2 = 
\sum_{\ell_1=\ell_\mathrm{min}}^\ell
\sum_{\ell_2=\ell_\mathrm{min}}^\ell
\sum_{\ell_3=\ell_\mathrm{min}}^\ell
\frac{\left[\alpha B_\kappa^w(\ell_1,\ell_2,\ell_3) - B^t_\kappa(\ell_1,\ell_2,\ell_3)\right]^2}{\mathrm{cov}(\ell_1,\ell_2,\ell_3)}
\end{equation}
and yields the best fitting $\alpha$ from the minimisation $\partial\chi^2/\partial\alpha = 0$. 

\begin{figure}
\begin{center}
\resizebox{\hsize}{!}{\includegraphics{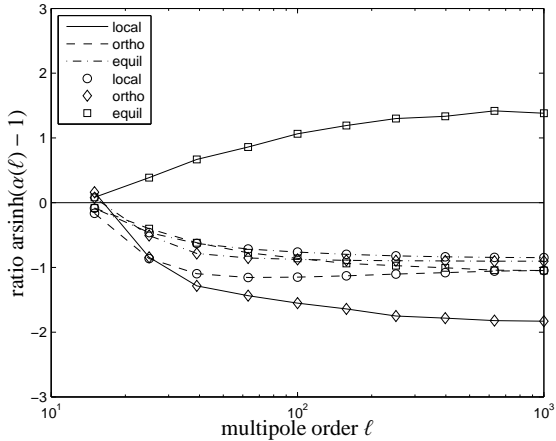}}
\end{center}
\caption{The ratio $\alpha(\ell)$ of inferred $f_\mathrm{NL}$-value to the true $f_\mathrm{NL}$-value as a function of maximum multipole order $\ell$. The true non-Gaussianity model is indicated by the line style, whereas the wrongly chosen non-Gaussianity model is given by the marker style: local (circles, solid lines), orthogonal (lozenges, dashed lines) and equilateral (squares, dash-dotted lines) non-Gaussianities.}
\label{fig_fnlratio}
\end{figure}

The variable $\alpha$ measures the ratio between the wrongly inferred non-Gaussianity parameter $f_\mathrm{NL}$ and the true value, and is given by Fig.~\ref{fig_fnlratio} as a function of maximum multipole order considered in the integration of the $\chi^2$-functional. For weak signals this ratio is very close to unity, and differences emerge when the integration is carried out to larger multipoles, and the signal becomes stronger. Misestimations in $f_\mathrm{NL}$ up to half an order of magnitude appear possible, including wrong signs for $f_\mathrm{NL}$-estimates. Most combinations of $B_\kappa^t$ and $B_\kappa^w$ yield very small values for the estimated $f_\mathrm{NL}$-parameter (equivalently, $\alpha\simeq-1$) when choosing the wrong non-Gaussianity. Again, the evaluations necessary for determining $\alpha$ are carried out as an MC-integration with the CUBA-library \citep{2005CoPhC.168...78H}.

\subsection{Would one notice fitting the wrong bispectrum?}
\begin{figure}
\begin{center}
\resizebox{\hsize}{!}{\includegraphics{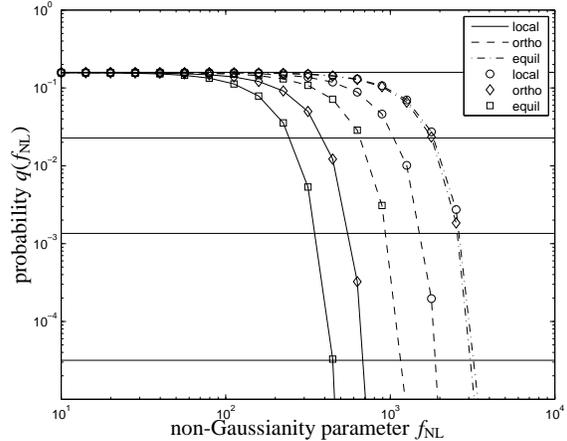}}
\end{center}
\caption{Probability $q(f_\mathrm{NL})$ of obtaining data as extreme as a wrong fit to the weak lensing bispectrum by choosing the wrong non-Gaussianity model (indicated by the line style) to data (indicated by the marker style): local (circles, solid lines), orthogonal (lozenges, dashed lines) and equilateral (squares, dash-dotted lines) non-Gaussianities. The horizontal lines indicate $1,2,3,4\sigma$ confidence intervals and $\ell$ has been set to $1000$.} 
\label{fig_pvalue}
\end{figure}

Now, the question appears if one would notice the assumption of a wrong primordial bispectrum when fitting for the non-Gaussianity parameter $f_\mathrm{NL}$. This can be quantified by the probability $q$ of obtaining data more extreme than the one at hand. This probability $q$ (Fisher's $p$-value) is given as a function of the true $f_\mathrm{NL}$ under the assumption of a Gaussian likelihood, which is well justified given the very large number of degrees of freedom \citep[although doubts have been raised about how accurate this is, see ][]{2011arXiv1104.0930S}. As shown in Fig.~\ref{fig_pvalue} for all combinations between true non-Gaussianity types and wrongly fitted non-Gaussianity models, this probability drops very rapidly towards very small numbers for $f_\mathrm{NL}$-values of a few hundred, indicating that it would be very difficult to reconcile non-Gaussianities of that strength with observations if the wrong non-Gaussianity model had been chosen. For $f_\mathrm{NL}$-values smaller than 100 the signal is so weak that no significant discrepancies between data and model appear, for any type of non-Gaussianity.

\subsection{Do parameter constraints depend on non-Gaussianity?}
\begin{figure}
\begin{center}
\resizebox{\hsize}{!}{\includegraphics{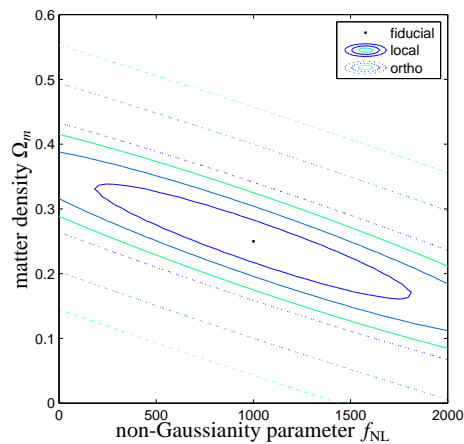}}
\end{center}
\caption{$1\sigma$, $2\sigma$ and $3\sigma$-constraints on the non-Gaussianity parameter $f_\mathrm{NL}$ and the matter density $\Omega_m$, for the local non-Gaussianity (solid lines) and the orthogonal non-Gaussianity (dotted lines) at $\ell=1000$, from a numerical MC-integration of the Fisher-matrix.}
\label{fig_constraint}
\end{figure}

The Fisher-matrix formalism \citep{1997ApJ...480...22T} is widely used in cosmology for deriving parameter forecasts, and requires in the case of the bispectrum as the signal-to-noise ratio the summation over all triangle configurations:
\begin{equation}
F_{\mu\nu} = 
\sum_{\ell_1=\ell_\mathrm{min}}^\ell
\sum_{\ell_2=\ell_\mathrm{min}}^\ell
\sum_{\ell_3=\ell_\mathrm{min}}^\ell 
\frac{\partial B_\kappa}{\partial x_\mu} 
\frac{1}{\mathrm{cov}(\ell_1,\ell_2,\ell_3)}
\frac{\partial B_\kappa}{\partial x_\nu}.
\end{equation}
The explicit summation can be replaced by a $\dd^3\ell$-integration, which can be carried out using the MC-technique outlined in Sect.~\ref{sect_s2n}. Resulting simultaneous constraints on $\Omega_m$ and $f_\mathrm{NL}$ from the weak shear bispectrum sourced only by primordial non-Gaussianities with no other priors are given in Fig.~\ref{fig_constraint} for local and orthogonal models at a reference $f_\mathrm{NL}=1000$. The equilateral bispectrum does not constrain the parameter pair in a meaningful way due to the weak signal.

\section{systematics due to structure formation}\label{sect_systematics}

\subsection{Can one subtract the structure formation bispectrum?}
Naturally, the small primordial non-Gaussianities are superseded by much stronger non-Gaussianities due to nonlinearities in the cosmic structure formation processes, which affects the measurability of $f_\mathrm{NL}$. There exists an accurate description of the structure formation bispectrum provided by Eulerian perturbation theory on the scales of interest \citep[compare Sect.~\ref{sect_sf_bispectrum}, and][]{2002PhR...367....1B}, if the cosmology is known -- but there are always uncertainties in the cosmological parameter set, which would result in an uncertainty in predicting the structure formation bispectrum. If the structure formation bispectrum is not properly subtracted from the observed bispectrum, there will be errors in the estimation of the non-Gaussianity parameter $f_\mathrm{NL}$ for the primordial bispectrum, which can be quantified with the $\chi^2$-functional,
\begin{equation}
\chi^2 =
\sum_{\ell_1=\ell_\mathrm{min}}^\ell
\sum_{\ell_2=\ell_\mathrm{min}}^\ell
\sum_{\ell_3=\ell_\mathrm{min}}^\ell
\frac{\left[\alpha B^t_\kappa(\ell_1,\ell_2,\ell_3) - \Delta B_\kappa(\ell_1,\ell_2,\ell_3)\right]^2}{\mathrm{cov}(\ell_1,\ell_2,\ell_3)},
\end{equation}
describing the fit of a primordial bispectrum $B^t_\kappa$ to data $\Delta B_\kappa$,
\begin{equation}
\Delta B_\kappa(\ell_1,\ell_2,\ell_3) = 
B^t_\kappa(\ell_1,\ell_2,\ell_3) + 
B^{t,\mathrm{SF}}_\kappa(\ell_1,\ell_2,\ell_3) - B^{w,\mathrm{SF}}_\kappa(\ell_1,\ell_2,\ell_3)
\end{equation}
which contain the true primordial bispectrum $B^t_\kappa$ itself, the very large structure formation bispectrum $B^{t,\mathrm{SF}}$ for the true cosmology, from which the structure formation bispectrum $B^{w,\mathrm{SF}}$ has been subtracted, possibly incompletely, by assuming the wrong cosmology. Derivation $\partial\chi^2/\partial\alpha = 0$ yields the best fitting $\alpha$, which is related to the misestimated $f_\mathrm{NL}^w = \alpha f_\mathrm{NL}$ and the deviation from the true non-Gaussianity $\delta = \alpha-1$.

\begin{figure}
\resizebox{\hsize}{!}{\includegraphics{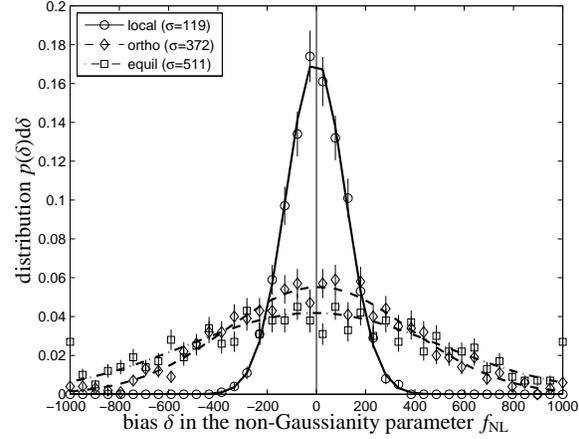}}
\caption{Distributions $p(\delta)\dd\delta$ of the bias $\delta$ between the inferred non-Gaussianity parameter and the true parameter $f_\mathrm{NL}$ if the structure formation bispectrum is not completely removed, due to uncertainties in the cosmological model, for local (circles, solid line), orthogonal (lozenges, dashed line) and equilateral (squares, dash-dotted line) non-Gaussianities. As priors, Euclid weak lensing and baryon acoustic oscillations were used.}
\label{fig_distribution}
\end{figure}

Distributions $p(\delta)\dd\delta$ have been derived for all bispectrum types by drawing $10^3$ samples from a Gaussian likelihood for the parameters $\Omega_m$, $\sigma_8$, $h$, $n_s$ and $w$ of a standard spatially flat dark energy model. The covariance matrix has been constructed using the icosmo resource for the Euclid weak lensing and BAO data \citep{2011A&A...528A..33R}, and provides an excellent prior on the cosmological parameters. By this sampling process of $\delta$ it is possible to propagate the entire uncertainty in the cosmological parameter set onto the estimate of $f_\mathrm{NL}$. As shown by Fig.~\ref{fig_distribution},  the resulting distribution is very close to Gaussian, with zero mean and standard deviations of $\sigma_{f_\mathrm{NL}}=119$ (local), $\sigma_{f_\mathrm{NL}}=372$ (orthogonal) and $\sigma_{f_\mathrm{NL}}=511$ (equilateral), which is similar to the statistical uncertainty of measuring $f_\mathrm{NL}$ and thus constitutes a serious error. The width of the distributions are independent of the true value $f_\mathrm{NL}$, and the relative error $\delta/f_\mathrm{NL}$ scales $\propto 1/f_\mathrm{NL}$. Misestimates of that magnitude make it very difficult to assign a primordial origin to a non-zero residual bispectrum, given the current bounds on $f_\mathrm{NL}$. All integrations were computed up to $\ell=1000$.

\subsection{What happens if a better prior is available?}
\begin{figure}
\resizebox{\hsize}{!}{\includegraphics{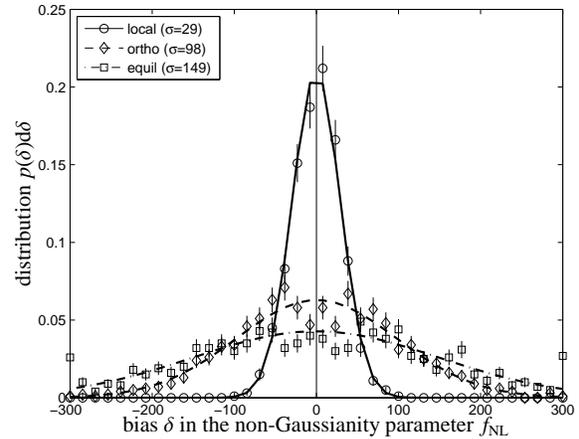}}
\caption{Distributions $p(\delta)\dd\delta$ of the bias $\delta$ with an enhanced prior for the cosmological model with constraints from weak lensing, baryon acoustic oscillations (both Euclid) and CMB temperature and polarisation spectra (Planck), for local (circles, solid line), orthogonal (lozenges, dashed line) and equilateral (squares, dash-dotted line) non-Gaussianities.}
\label{fig_distribution_enhanced}
\end{figure}

The uncertainty in predicting the weak shear bispectrum generated by nonlinear structure formation can be reduced if a stronger prior on the cosmological parameter set is available or if the complexity of the model is reduced, e.g. if the $w$CDM dark energy cosmology would be replaced by the simpler $\Lambda$CDM cosmology. Fig.~\ref{fig_distribution_enhanced} illustrates the distributions $p(\delta)\dd\delta$ of the difference $\delta$ between the inferred non-Gaussianity parameter and the true parameter, if the contamination of the bispectrum is computed by drawing $10^3$ sample $w$CDM cosmologies from a Gaussian likelihood whose covariance matrix incorporates constraints from Euclid weak shear spectra, Euclid baryon acoustic oscillations and in addition Planck's constraints from the observation of primary CMB temperature and polarisation spectra. In comparison to the distributions shown in Fig.~\ref{fig_distribution}, the width is now much reduced, by about a factor of 4, allowing measurements down to smaller values for $f_\mathrm{NL}$. The specific uncertainties are $\sigma_{f_\mathrm{NL}}=29$ (local), $\sigma_{f_\mathrm{NL}}=98$ (orthogonal) and $\sigma_{f_\mathrm{NL}}=149$ (equilateral), all stated as standard deviations of the distribution $p(\delta)\dd\delta$. These uncertainties are below current bounds on $f_\mathrm{NL}$ and are small enough for studies of primordial non-Gaussianities. Again, all integrations were carried out up to $\ell=1000$.

\section{Summary}\label{sect_summary}
The topic of this paper are measurements of primordial bispectra in weak shear data from Euclid, comparisons between different types of non-Gaussianity configuration dependences, statistical questions concerning the inference of the non-Gaussianity parameter $f_\mathrm{NL}$ and the removal of the much stronger structure formation induced bispectrum. Although not as sensitive as observations of the CMB-bispectrum or the galaxy bispectrum for scale-free non-Gaussianities, weak lensing can place useful independent constraints on non-Gaussianities, in particular on smaller scales where CMB bounds might not apply. It is less prone to systematics than other large-scale structure probes and provides a direct linear mapping of the density field, which conserves its statistical properties.

\begin{enumerate}
\item{Primordial non-Gaussianities provide a rather weak signal in the weak shear bispectrum \citep[because of the Gaussianising effect of the line-of-sight integration, see][]{2011arXiv1104.0926J}, and signal-to-noise ratios of order unity can only be expected for $f_\mathrm{NL} = 200, 575, 1628$ for local, orthogonal and equilateral non-Gaussianities, respectively, where this measurement is most sensitive to scales $0.1\ldots1~(\mathrm{Mpc}/h)^{-1}$. These bounds are weaker than those from e.g. observations of the CMB bispectrum, but will serve nevertheless for cross validation, in particular given the absence of strong systematics is weak shear data, or as bounds on scale-dependent non-Gaussianity, because weak lensing maps out scales which are not constrained by the CMB and is sensitive to scales probed by number counts.}
\item{Configuration space integrations can be very efficiently carried out by Monte-Carlo integration schemes, at a fraction of the computational cost. Computations of the signal-to-noise ratio, of $\chi^2$-functionals or of the Fisher-matrix $F_{\mu\nu}$ can be done with accuracies below a percent with $\mathcal{O}(10^5)$ evaluations instead of $\mathcal{O}(10^9)$ evaluations for the direct sum over $\ell_1$, $\ell_2$ and $\ell_3$. Very good results were obtained with the CUBA library \citep{2005CoPhC.168...78H}.}
\item{Fitting the wrong bispectrum type to data yields serious misestimates in the non-Gaussianity parameter $f_\mathrm{NL}$. Depending on the combination of true and false model there are two cases: either the estimated $f_\mathrm{NL}$ becomes very small, or the estimate for $f_\mathrm{NL}$ is a factor of $\sim\pm 3$ too large. When looking at numerical values for the $\chi^2$-functional, one would notice strong discrepancies between data and model when fitting the wrong non-Gaussianity type from values of $f_\mathrm{NL}$ of a few hundred on.}
\item{The much stronger structure formation bispectrum can be subtracted with a prediction of its bispectrum from perturbation theory if the cosmology is known precisely enough. Propagating the uncertainty in the cosmological parameter set onto the misestimation of $f_\mathrm{NL}$ if the structure formation bispectrum is not correctly subtracted yielded typical uncertainties of 29, 98 and 149 for local, orthogonal and equilateral non-Gaussianities, much less than the statistical accuracy. As a prior on the cosmological parameters we assumed a Gaussian likelihood for a $w$CDM model combining Euclid's weak shear with baryon acoustic oscillations and Planck's observations of primary CMB anisotropies. Similar ratios between the numerical value of $f_\mathrm{NL}$ and the standard cosmological parameters were found by \citet{2011MNRAS.411..595P}.}
\end{enumerate}
Many of our investigations can be straightforwardly generalised to other probes of large-scale structure statistics. We intend to generalise our investigations to higher polyspectra and to apply ideas from Bayesian model selection \citep{2007MNRAS.378...72T,2008ConPh..49...71T} for assigning probabilities to the problem of choosing the correct non-Gaussianity type.

\section*{Acknowledgements}
We acknowledge the use of the formidable icosmo resource for computing a joint Fisher matrix for baryon acoustic oscillations and the weak shear spectrum. We would like to thank Maik Weber and Youness Ayaita for providing a CMB Fisher matrix for the Planck survey characteristics, Matthias Bartelmann for his suggestions, Patricio Vielva for his help in clarifying statistical questions and Tommaso Giannantonio for thoughtful comments on our draft. Our work was supported by the German Research Foundation (DFG) within the framework of the excellence initiative through the Heidelberg Graduate School of Fundamental Physics and by the German National Academic Foundation. This paper is listed as preprint BI-TP 2011/20.

\bibliography{bibtex/aamnem,bibtex/references}
\bibliographystyle{mn2e}

\appendix

\section{configuration dependence}\label{sect_configuration}
\begin{figure*}
\begin{center}
\begin{tabular}{cc}
\resizebox{0.48\hsize}{!}{\includegraphics{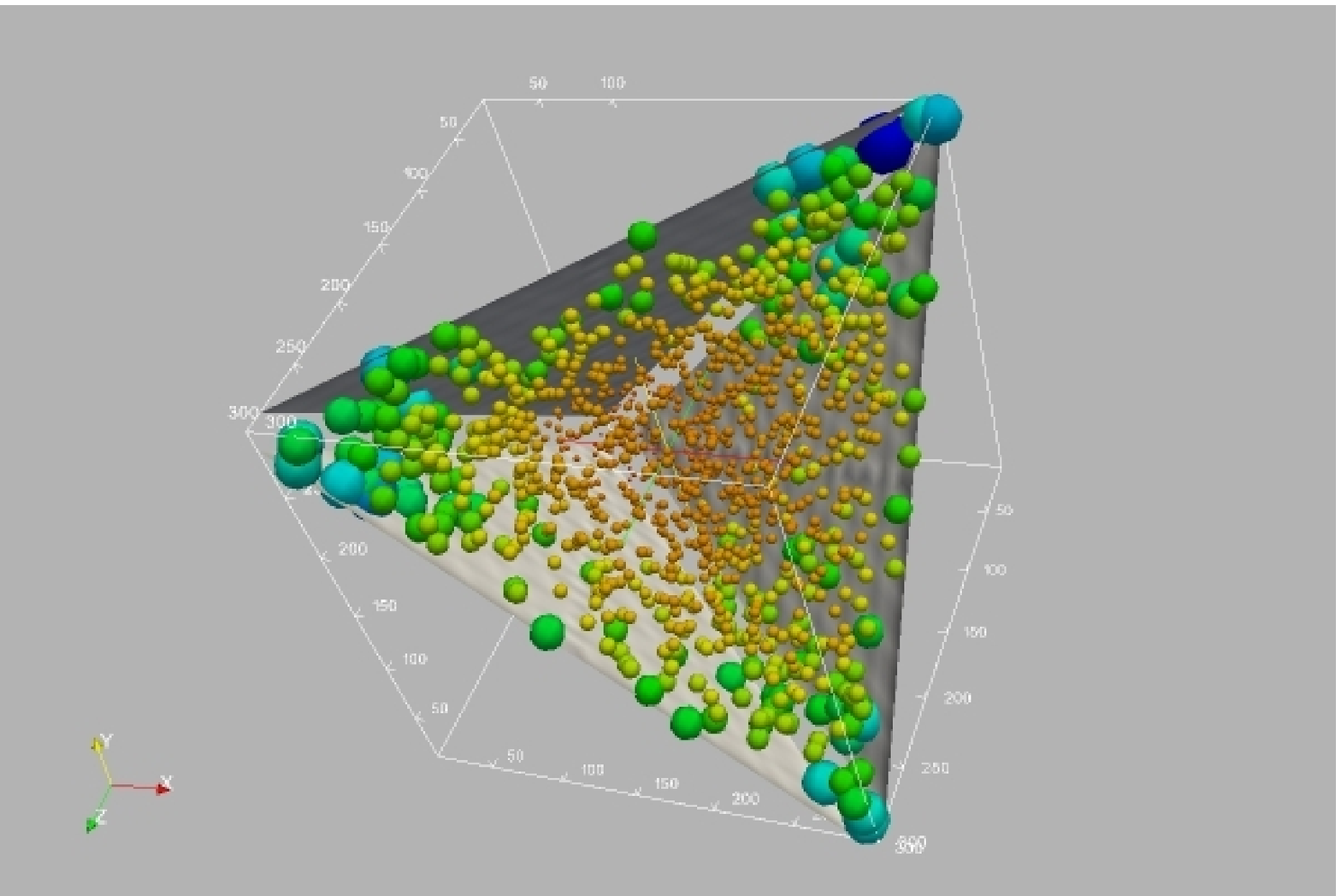}} &
\resizebox{0.48\hsize}{!}{\includegraphics{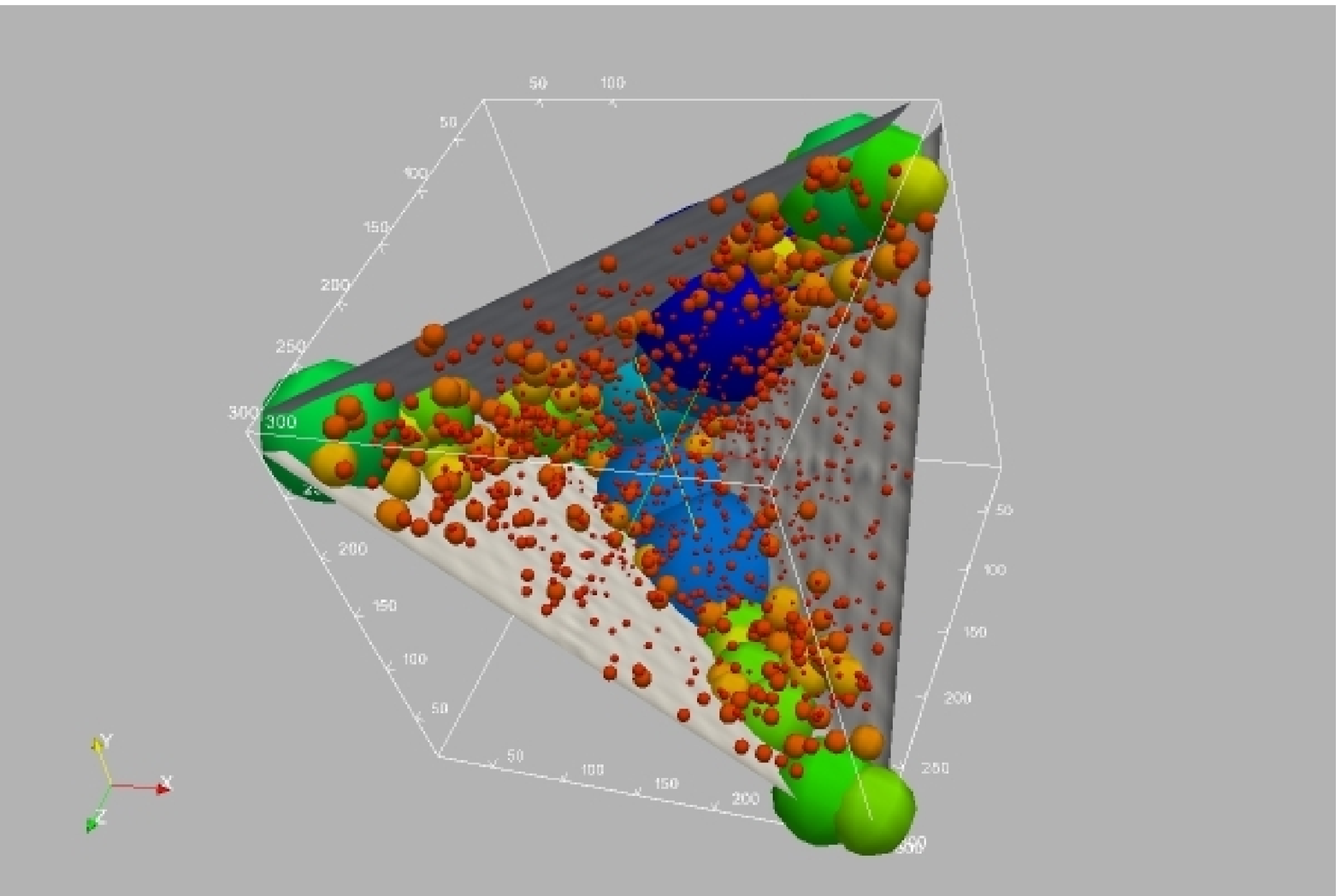}} \\
\resizebox{0.48\hsize}{!}{\includegraphics{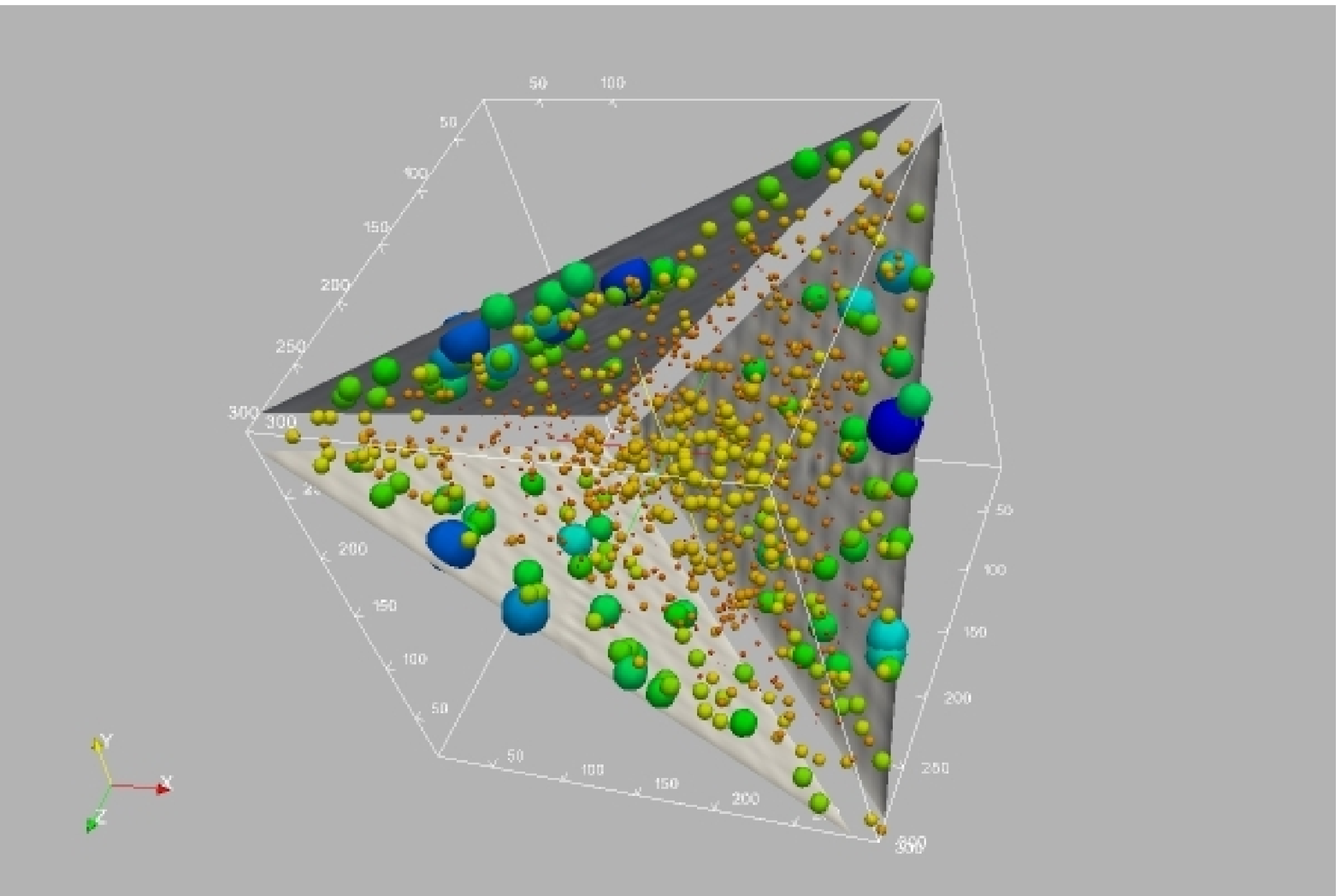}} &
\resizebox{0.48\hsize}{!}{\includegraphics{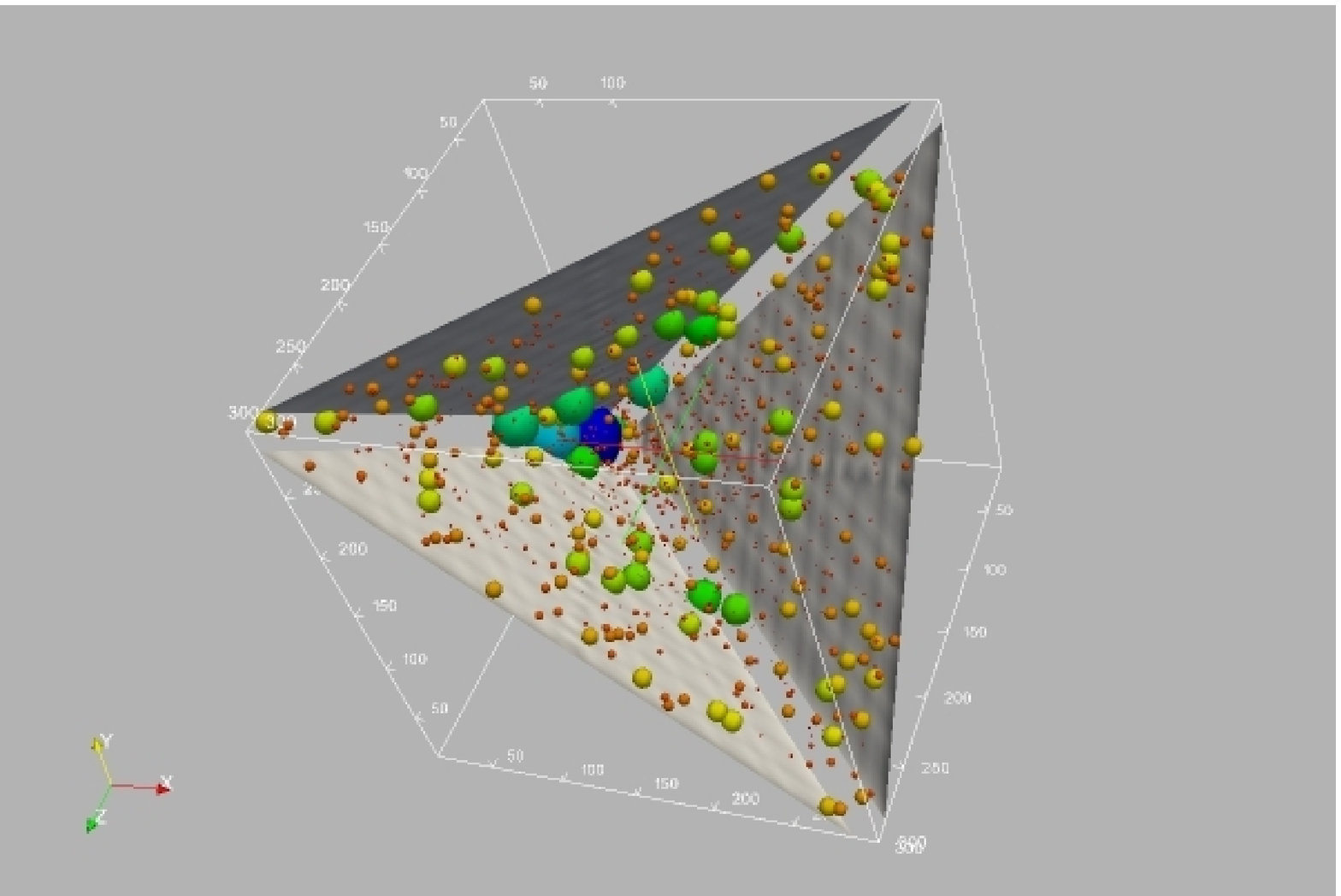}} \\
\resizebox{0.48\hsize}{!}{\includegraphics{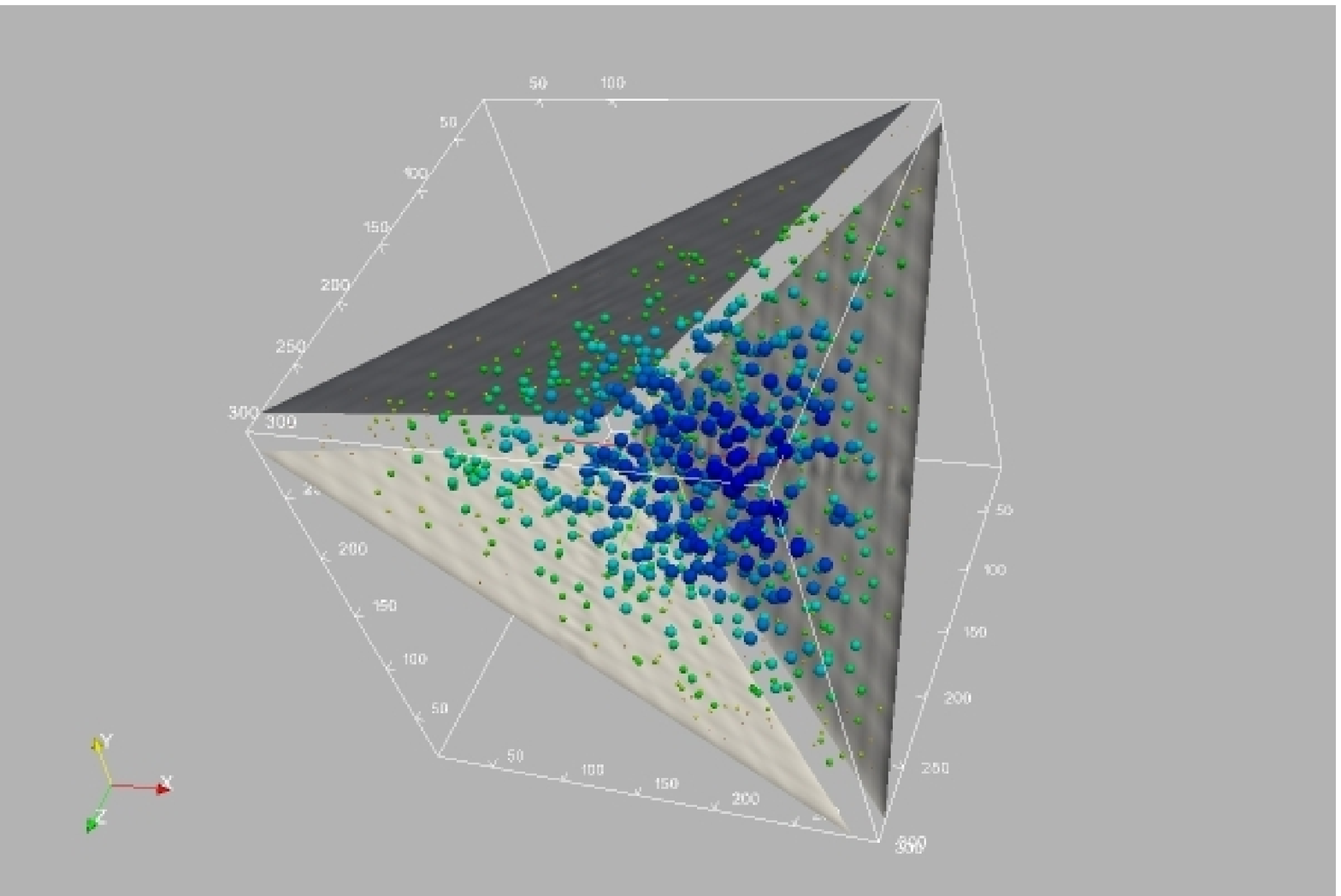}} &
\resizebox{0.48\hsize}{!}{\includegraphics{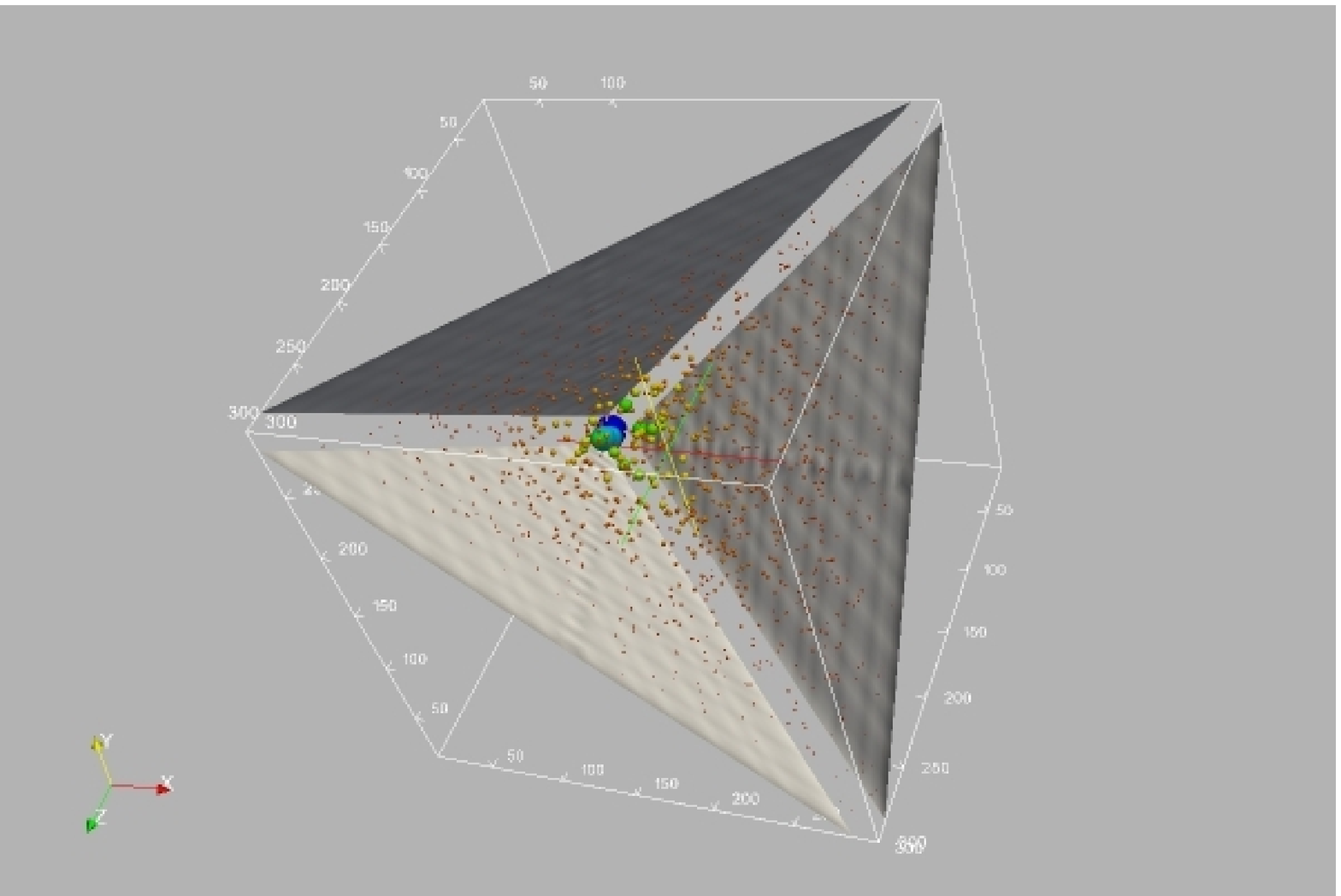}}
\end{tabular}
\end{center}
\caption{Configuration dependence $(\ell_1\ell_2\ell_3)^{4/3}B_\kappa(\ell_1,\ell_2,\ell_3)$ (first column) and signal-to-noise ratio $B_\kappa(\ell_1,\ell_2,\ell_3)/\sqrt{\mathrm{cov}(\ell_1,\ell_2,\ell_3)}$ (second column) of the weak lensing bispectrum, for local (first row), orthogonal (second row) and equilateral (third row) non-Gaussianities. The size of the blobs and their colour is proportional to the bispectrum, where a correct relative normalisation in the columns is maintained. Configurations outside the grey bounding planes violate the triangle inequality.}
\label{fig_config}
\end{figure*}

Fig.~\ref{fig_config} compares the configuration dependence of the bispectrum and of the signal strength in a weak lensing experiment. As a representation, we chose to plot the dimensionless weak convergence bispectrum $(\ell_1\ell_2\ell_3)^{4/3}B_\kappa(\ell_1,\ell_2,\ell_3)$ and the convergence bispectrum in units of the noise, $B_\kappa(\ell_1,\ell_2,\ell_3)/\sqrt{\mathrm{cov}(\ell_1,\ell_2,\ell_3)}$, which when added in quadrature yields the signal-to-noise ratio. The factor $(\ell_1\ell_2\ell_3)^{4/3}\sim\ell^4$ makes the angular bispectra dimensionless.

\bsp

\label{lastpage}

\end{document}